\documentclass[epj]{svjourmod}

\usepackage{amsmath,amssymb,amsfonts}
\usepackage{graphicx}
\usepackage{citesort}

\newcommand{\beq}{\begin{equation}}
\newcommand{\eeq}{\end{equation}}
\newcommand{\Order}{\mathcal{O}}
\newcommand{\Lagr}{\mathcal{L}}
\newcommand{\M}{\mathcal{M}}
\newcommand{\sh}{\!\not\!}
\newcommand{\mpi}{M_\pi^2}
\newcommand{\mk}{M_K^2}
\newcommand{\Em}{E_{\rm max}}
\newcommand{\Ec}{E_{\rm cut}}
\newcommand{\nn}{\nonumber\\}
\DeclareMathOperator{\Li}{Li}
\renewcommand{\vec}[1]{\mathbf{#1}}

\begin{document}

\title{\\[-0.95cm]\phantom{ }\hfill{\scriptsize \textnormal{HISKP--TH--10/17}}\\[1mm]
Radiative corrections in \boldmath{$K\to\pi\ell^+\ell^-$} decays}
\titlerunning{Radiative corrections in $K\to\pi\ell^+\ell^-$ decays}

\author{Bastian Kubis, Rebekka Schmidt}

\institute{
   Helmholtz-Institut f\"ur Strahlen- und Kernphysik (Theorie)
   and 
   Bethe Center for Theoretical Physics,
   Universit\"at Bonn, \linebreak D--53115~Bonn, Germany
}

\authorrunning{B. Kubis and R. Schmidt}

\date{
}

\abstract{
We calculate radiative corrections to the flavor-changing neutral current
process $K\to\pi\ell^+\ell^-$, both for charged and neutral kaon decays.
While the soft-photon approximation is shown to work well for the muon channels, 
we discuss the necessity of further phase space cuts with electrons in the final state.
It is also shown how to transfer our results to other decays such as 
$\eta\to\gamma\ell^+\ell^-$ or $\omega\to\pi^0\ell^+\ell^-$.
}

\maketitle

\section{Introduction}

The radiative non-leptonic decay $K\to\pi \ell^+\ell^-$ ($\ell=e,\,\mu$) offers insight 
into the nature of the weak interactions at low energies.
These flavor-changing neutral current transitions are suppressed to the
one-loop level in the standard model, where the CP-allowed channels 
$K^\pm \to \pi^\pm \ell^+\ell^-$ and
$K_S \to \pi^0 \ell^+\ell^-$ are expected to be dominated by
one-photon exchange ($K\to\pi\gamma^*$).  
The real-photon decay $K\to\pi\gamma$ is forbidden by gauge invariance, 
and in the effective-theory description of chiral perturbation theory,
also the virtual transition is suppressed to the one-loop level~\cite{Ecker:1987qi}.
$K\to\pi \ell^+\ell^-$ has since been investigated in chiral perturbation theory
beyond one-loop order~\cite{D'Ambrosio:1998yj,Friot:2004yr} and 
with isospin breaking included~\cite{Bijnens:2007xa}.

The least rare of these decays, $K^\pm \to \pi^\pm e^+e^-$, 
the first to be discovered at the CERN PS~\cite{Bloch:1974ua} 
and subsequently studied with increasing precision at BNL~\cite{Alliegro:1992pp},
has now been investigated by the modern high-precision experiments E865 in Brookhaven~\cite{Appel:1999yq}
and NA48/2 at CERN~\cite{Batley:2009pv}, both with numbers of events of the order of $10^4$.
For the muon decay channel $K^\pm \to \pi^\pm \mu^+\mu^-$,
a few hundred events have been recorded at BNL~\cite{Adler:1997zk,Ma:1999uj}
and Fermilab~\cite{Park:2001cv},
and NA48/2 is in the process of analyzing a sample of $\sim 3000$ events~\cite{Goudzovski:2009mg};
the $K_S$ decays have been seen by NA48/1,
though with a handful of events only so far~\cite{Batley:2003mu,Batley:2004wg}.
As a welcome side effect of the upcoming NA62 experiment dedicated to the search 
of the very rare $K\to\pi\nu\bar\nu$ decays, a strong future increase in statistics
for $K\to \pi\ell^+\ell^-$ is also expected, 
making the availability of reliable (electromagnetic) radiative corrections 
an indispensable requirement.

As the chiral representation of the $K\to\pi\gamma^*$ form factor starts at
the one-loop level, a full analysis of radiative corrections in the framework
of chiral perturbation theory seems highly impractical, as it would require
a two-loop calculation already at leading non-trivial order.  
Furthermore, the counterterms appearing in such processes (weak non-leptonic
with electromagnetic corrections at two loops) have never been considered,
let alone the finite parts of such counterterms estimated phenomenologically.
We therefore follow a more pragmatic approach akin to the study of 
soft-photon corrections in meson decays in Ref.~\cite{Isidori}
and restrict ourselves largely to universal radiative corrections, assuming
a linear form factor; in this way, a technically simpler one-loop calculation suffices.
Given that the data are still compatible with such a linear form factor approximation~\cite{Appel:1999yq,Batley:2009pv},
this seems a reasonable procedure.  Furthermore, as the only non-analyticity in the form factor 
occurring within the physical decay region is the $\pi^+\pi^-$ intermediate state
that necessarily has to appear in a P-wave~\cite{D'Ambrosio:1998yj},
the deviation from an analytic behavior is moderate, with the corresponding
imaginary part rising slowly above the $\pi\pi$ threshold.  

We shall see below, after introducing the necessary formalism in Sect.~\ref{sec:form},
that for the most interesting corrections concerning
the invariant mass spectrum of the lepton--antilepton pair,
the assumption of a specific form factor is actually not of high concern, 
as it factorizes neatly.  
These results are contained in Sect.~\ref{sec:dGds}.
We discuss the soft-photon approximation for the bremsstrahlung, and
how the appearance of collinear singularities for the electron--positron
final state enforces to go beyond that, introducing additional phase space cuts.
The form factor slope only affects those radiative corrections that
induce an $\ell^+\ell^-$ asymmetry, which we investigate in Sect.~\ref{sec:asym}.
Despite quite different hadronic currents contracted with the leptonic ones, 
we  demonstrate in Sect.~\ref{sec:other} that a significant part of our results
can be transferred directly to other meson decays involving lepton--antilepton pairs in the final state,
such as the Dalitz decays $\pi^0\to\gamma e^+e^-$, $\eta\to\gamma\ell^+\ell^-$, or the vector meson
conversion decays $\omega\to\pi^0\ell^+\ell^-$, $\phi\to\eta\ell^+\ell^-$ etc.
Finally, we conclude with a summary;
several of the more laborious formulas are relegated to the appendices.

\section{Formalism}\label{sec:form}

\begin{sloppypar}
We consider the decay $K(k)\to\pi(p)\ell^+(p_+)\ell^-(p_-)$ in terms of the kinematic variables
$s=(k-p)^2$, $t=(k-p_+)^2$, $u=(k-p_-)^2$.
For the CP-allowed ($K^\pm$ and $K_S$) channels, the $K\to\pi\gamma^*$ form factor $F(s)$ is defined as
\begin{align}
&i\int d^4x \,e^{i(k-p)x}\langle \pi(p)|T\{J^\mu_\text{em}(x)\Lagr_{\Delta S=1}(0)\}|K(k)\rangle \nn
&=\frac{e\,F(s)}{M_K^2(4\pi)^2}\left[s(k+p)^\mu- (k^2-p^2)(k-p)^\mu\right] ~,\label{eq:FF}
\end{align}
which on the kaon and pion mass shell simplifies according to $k^2=M_K^2$, $p^2=M_\pi^2$.
Here, $J^\mu_\text{em}$ is the electromagnetic current and $\mathcal{L}_{\Delta S=1}$ the strangeness-changing
non-leptonic weak Lagrangian.  
The form factors $F(s)$ are (potentially) very different for $K^\pm$ and $K_S$ decays~\cite{D'Ambrosio:1998yj};
for simplicity we refrain from reflecting this explicitly in the notation.
Contracting Eq.~\eqref{eq:FF} with the leptonic current
yields the decay amplitude
\beq
\M(K\to\pi\ell^+l^-) = -\frac{e^2\,F(s)}{M_K^2(4\pi)^2}\bar{u}(p_-)(\sh k+\sh p) v(p_+) ~,\label{eq:A0}
\eeq
which, when squared and summed over fermion spins, results in the Dalitz plot distribution
\beq
d\Gamma = \frac{\alpha^2|F(s)|^2}{4(4\pi)^5M_K^7} \left(\lambda-\nu^2\right)ds\,d\nu ~,\label{eq:Dalitz}
\eeq
where $\alpha = e^2/4\pi$ is the fine structure constant, $\lambda \doteq \lambda(\mk,s,\mpi)$ with the 
K\"all\'en function $\lambda(a,b,c)=a^2+b^2+c^2-2(ab+ac+bc)$, and $\nu=t-u$. 
Integrating over $\nu$, with the limits of the Dalitz plot given by 
\beq
-\nu_{\rm max} \leq \nu \leq \nu_{\rm max} ~,~~ 
\nu_{\rm max} = \sigma \lambda^{1/2} ~,~~ 
\sigma = \sqrt{1-\frac{4m^2}{s}} ~,
\eeq
and $m$ is the mass of the lepton involved,
we find the spectrum with respect to the dilepton invariant mass as
\beq
\frac{d\Gamma}{ds} = \frac{\alpha^2|F(s)|^2}{2(4\pi)^5M_K^7} \lambda^{3/2}\sigma\Big(1-\frac{\sigma^2}{3}\Big) ~.
\eeq
\end{sloppypar}

\begin{figure}
\centering
\includegraphics[width=0.95\linewidth]{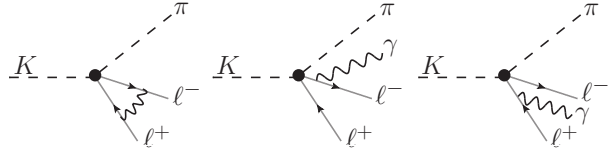}
\caption{Feynman graphs contributing to radiative corrections in 
$K_S\to\pi^0\ell^+\ell^-$ through virtual- and real-photon processes.
We do not show diagrams responsible for (electromagnetic) wave function renormalization.}
\label{fig:feynman1}
\end{figure}

Due to the Dirac structures in the leptonic current, radiative corrections in these decays
cannot easily be written as one overall correction factor to the Dalitz plot
distribution Eq.~\eqref{eq:Dalitz} (compare Ref.~\cite{Isidori}).
This is easily seen by the virtual-photon exchange between lepton and antilepton,
see Fig.~\ref{fig:feynman1},
which is precisely the diagram responsible for the generation of the anomalous magnetic
moment at leading order, and hence a new Dirac structure.  
In general, we write the radiative corrections to the Dalitz plot in the form
\beq
d\Gamma = \frac{\alpha^2|F(s)|^2}{4(4\pi)^5M_K^7} 
\Big[\big(\lambda-\nu^2\big)(1+\omega+\hat\omega) + \nu^2\bar\omega + m^2\nu\tilde\omega \Big]ds\,d\nu ~,\label{eq:DalitzEM}
\eeq
where $\omega$, $\bar\omega$, and $\tilde\omega$ are even functions in $\nu$, while $\hat\omega$ is odd;
all four of them are of order $\alpha$.
This leads to a corrected dilepton spectrum according to
\begin{align}
\frac{d\Gamma}{ds} &= \frac{\alpha^2|F(s)|^2}{2(4\pi)^5M_K^7} \lambda^{3/2}\sigma\Big(1-\frac{\sigma^2}{3}\Big)
(1+\Omega ) ~, \nn
\Omega &= \omega+\frac{\sigma^2}{3-\sigma^2}\bar\omega ~.\label{eq:Omegadef}
\end{align}
The relation between $\Omega$, $\omega$, and $\bar\omega$ in Eq.~\eqref{eq:Omegadef} will
be implicitly assumed for all parts and approximations discussed for these correction factors
in the following section.
The factors $\hat\omega$ and $\tilde\omega$ (which are only non-vanishing for the \emph{charged} kaon decay) 
cancel in $d\Gamma/ds$ upon integration of the terms odd in $\nu$.
They do, however, induce an asymmetry in $\nu$ (or $\ell^+\ell^-$ asymmetry) to which, 
in turn, $\omega$ and $\bar\omega$ do not contribute.  
This asymmetry will be defined and discussed in Sect.~\ref{sec:asym}.
Finally, we remark that, in this article, we do not discuss effects due to vacuum polarization in the 
one-photon exchange graphs, which can simply be taken care of by the properly renormalized
running fine structure constant $\alpha$.

\section{Radiative corrections for \boldmath{$d\Gamma/ds$}}\label{sec:dGds}

\subsection{\boldmath{$K\to\pi \mu^+\mu^-$}: soft-photon approximation}\label{sec:musoft}

It is well known that loop corrections involving massless particles (e.g., photons)
diverge in the infrared.  These divergences are exactly canceled by the contributions from
real-photon emission (bremsstrahlung)~\cite{BlochNordsieck,YennieFrautschi,Weinberg:1965nx}, 
which is typically included up to some maximum
photon energy $\Em$; remnants of the infrared divergence can then be seen 
by terms depending logarithmically on $\Em$.  
The finite, not logarithmically enhanced terms depend on the choice of reference frame
in which this cut is applied.  
While a conventional choice in decays is the rest frame of the decaying particle,
we would like to argue that this is not very advantageous in the case at hand:
as the predominant interest in this decay lies in the extraction of the form factor
$F(s)$ from data, a preferred option is to use a cutoff in the rest frame of the 
$\ell^+\ell^-\gamma$ system in the final state, or $\vec{k}-\vec{p}=0$. 
In this way, the bremsstrahlung contribution---like the rest of the electromagnetic
corrections---is essentially a function of $s$ only and does not induce any $\ell^+\ell^-$
asymmetry by itself.
The results we present in the following always refer to this choice for the photon-energy cut.
$\Em$ is then given by the minimum of a cutoff energy $\Ec$ and the kinematic limit,
\beq
\Em = {\rm min}\Big\{\Ec, \frac{s-4m^2}{2\sqrt{s}} \Big\} ~.
\eeq

\begin{sloppypar}
The bremsstrahlung calculation is typically the hardest part of the radiative corrections
at $\Order(\alpha)$.  It is drastically simplified in the so-called soft-photon approximation, 
which consists essentially in neglecting the photon momentum in the overall momentum-conserving
$\delta$-function; as a consequence, also all kinematic variables can be chosen identical
to the non-radiative process.  It is easy to show that the soft-photon approximation
differs from the exact results by terms of order $\Em$, i.e.\ 
reproduces the logarithmic terms as well as terms of order $\Em^0$ correctly.
With a typical cutoff of $\Ec = 20\ldots 30$~MeV, and a process 
such as $K\to\pi \mu^+\mu^-$ in which all mass scales
are significantly larger than $\Em$, the soft-photon approximation is therefore
expected to work very well.  We shall verify this below.

There may actually be other processes contributing to real-photon emission besides bremsstrahlung,
i.e.\ the radiation of a soft photon from one of the external charged particles, namely
so-called structure-dependent contributions.  These structure-dependent terms can be shown
to also be of $\Order(\Em)$, hence comparable to those neglected in the soft-photon approximation.
In the process at hand, there is a very large structure-dependent contribution 
from $K^+\to\pi^+\pi^0$ with subsequent Dalitz decay $\pi^0\to\gamma e^+e^-$,
which however occurs at fixed $s=M_{\pi^0}^2$.  
To avoid contamination with structure-dependent photon radiation, the cutoff $\Em$ may
not be chosen too large.  As a side remark, this should currently not be a major issue, 
as even the main (non-radiative) decay $K^+\to\pi^+e^+e^-$ is only measured for 
$s>M_{\pi^0}^2$ to avoid background from $K^+\to\pi^+\pi^0$ with subsequent $\pi^0\to e^+e^-$.

\begin{figure}
 \centering
 \includegraphics[width=\linewidth]{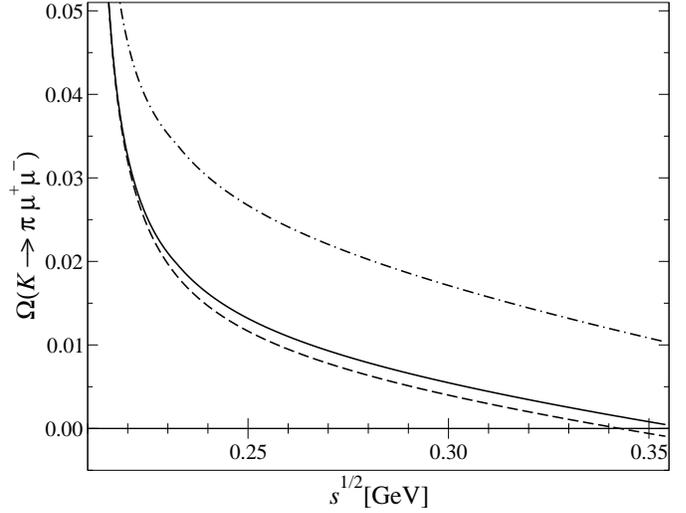}
 \caption{Correction factor for $d\Gamma/ds(K_S\to\pi^0 \mu^+\mu^-)$, both
in soft-photon approximation (full line) and for the exact result (dashed),
plus the correction factor for $d\Gamma/ds(K^+\to\pi^+ \mu^+\mu^-)$
in soft-photon approximation (dash-dotted curve).
All terms are given for a photon-energy cutoff $\Ec=20$~MeV.}
 \label{fig:Kmu}
\end{figure}

We begin by discussing the simplest of the decay channels with respect to 
radiative corrections, $K_S\to\pi^0 \mu^+\mu^-$.  Obviously, photons can only
couple to the charged leptons in the final state. The diagrams contributing
to both virtual- and real-photon corrections are shown in Fig.~\ref{fig:feynman1}.
The resulting correction factors are given by
$\omega(K_S\to\pi^0 \mu^+\mu^-) = \omega_{\ell^+\ell^-}^{\rm soft}$
and correspondingly for $\bar\omega$ and $\Omega$, where
\begin{align}
\omega_{\ell^+\ell^-}^{\rm soft} &= \frac{\alpha}{\pi}
\bigg\{2 \left(\frac{1 + \sigma^2}{2 \sigma}\log\frac{1+\sigma}{1-\sigma} - 1\right)
\left(\log\frac{2 \Em}{m} + 1\right)\nonumber\\ 
&+ \frac{1+\sigma^2}{2\sigma} 
\bigg[\pi^2 - 2 \Li\Big(\frac{2\sigma}{\sigma+1}\Big) + 2 \Li\Big(\frac{2\sigma}{\sigma-1}\Big)\bigg]\nonumber\\ 
&+ \frac{\sigma}{2}\log\frac{1+\sigma}{1-\sigma} - \frac{1}{2}  \bigg\} ~, \nn
\bar\omega_{\ell^+\ell^-}^{\rm soft} &= - \frac{\alpha}{\pi}\frac{1-\sigma^2}{2\sigma}\log\frac{1+\sigma}{1-\sigma} ~,
\label{eq:Ollsoft}
\end{align}
where $\Li(z)$ is the dilogarithm or Spence function,
\beq
\Li(z) = - \int_0^z\frac{\log (1-t)}{t}dt ~.
\eeq
The expressions in Eq.~\eqref{eq:Ollsoft} are ultraviolet finite and do not require a counterterm;
this is a consequence of the Ward identity in QED.
The resulting correction factor $\Omega_{\ell^+\ell^-}^{\rm soft}$ is plotted in Fig.~\ref{fig:Kmu}.
At small $s$, it is obviously dominated by the Coulomb pole in Eq.~\eqref{eq:Ollsoft},
\beq
\Omega_{\ell^+\ell^-}^{\rm C} = \frac{\alpha}{\pi}\frac{1+\sigma^2}{2\sigma} \pi^2~, \label{eq:Coulomb}
\eeq
which is the leading approximation (in $\alpha$) to the Gamow--Sommerfeld factor~\cite{Gamow,Sommerfeld}.
Next-to-leading order corrections (of $\Order(\alpha^2)$) become important as soon as 
$\alpha(1+\sigma^2)/2\sigma$ is not small any more, which, in the present case of muons involved,
happens in the sub-keV region above threshold.  Given the expected energy resolution of kaon decay experiments, 
the approximation Eq.~\eqref{eq:Coulomb} is therefore surely sufficient.

The exact expression $\Omega_{\ell^+\ell^-}$, without the soft-photon approximation, 
is given in Appendix~\ref{app:Ollexact}.  For illustration, we compare
exact and soft-photon form for $\Ec=20$~MeV in Fig.~\ref{fig:Kmu}: the differences are small, 
of the order of 0.2\% at most.  This is in accordance 
with a corresponding comparison for radiative corrections in $K\to 3\pi$ decays~\cite{Photons}
(which has rather similar kinematics to $K\to \pi \mu^+\mu^-$).
Changing the cutoff to $\Ec=30$~MeV hardly changes the picture.
\end{sloppypar}

\begin{figure}
\centering
\includegraphics[width=0.95\linewidth]{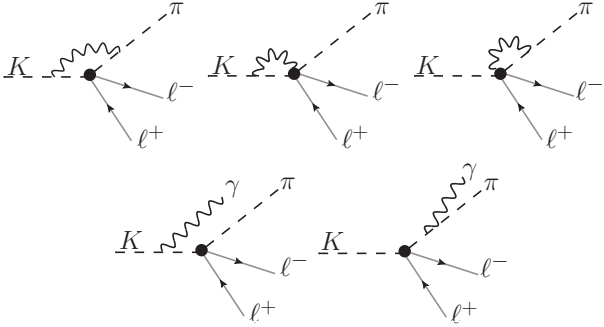}
\caption{Additional Feynman graphs contributing to radiative corrections in 
$d\Gamma/ds(K^+\to\pi^+\ell^+\ell^-)$ through virtual (top) and real (bottom) photon processes.
Again, diagrams contributing to wave function renormalization are not shown explicitly.}
\label{fig:feynman2}
\end{figure}

The generalization to the correction factor $\Omega(K^+\to\pi^+ \mu^+\mu^-)$ 
for the dilepton spectrum is surprisingly simple.  It is given as 
\beq
\Omega(K^+\to\pi^+ \mu^+\mu^-) = \Omega_{\ell^+\ell^-} + \Omega_{K\pi} ~,
\eeq
where $\Omega_{K\pi}$ is calculated exclusively from the diagrams displayed in Fig.~\ref{fig:feynman2},
i.e.\ those graphs that would also constitute the radiative corrections 
in $K^+\to\pi^+ \nu\bar\nu$, say.
All virtual-photon corrections linking mesons and leptons, or the interference of 
meson and lepton bremsstrahlung, are odd in $t-u$ and hence cancel in the dilepton spectrum
(they are part of the correction factors $\hat\omega$ and $\tilde\omega$ 
to be discussed in Sect.~\ref{sec:asym}).
Furthermore, $\Omega_{K\pi}=\omega_{K\pi}$, as obviously
radiative corrections only concerning the hadronic current do not induce a new Dirac structure,
$\bar\omega_{K\pi}=0$.
We only calculate $\omega_{K\pi}$ in the soft-photon approximation, for which we find the analytic
representation
\begin{align}
\omega_{K\pi}^{\rm soft}&= \frac{\alpha}{\pi}\bigg\{2 \left[\frac{1-\Sigma}{2\rho} 
\log\frac{\Delta^2 - (1 + \rho)^2}{\Delta^2 - (1 - \rho)^2} - 1\right] \nn
& \quad \times\left(\log\frac{2 \Em}{\sqrt{M_K M_\pi}} + 1\right) \nn
&+ \frac{\Delta}{2\rho}\log\frac{(\Delta + \rho)^2 - 1}{(\Delta - \rho)^2 - 1} 
+ \frac{1}{2\rho} \log\frac{\Delta^2 - (1 + \rho)^2}{\Delta^2 - (1 \ - \rho)^2} \nn
&+ \frac{1 - \Sigma}{4 \rho} \bigg[2\Li\Big( \frac{2 \rho}{\rho + \Delta - 1}\Big) 
+ 2\Li\Big(\frac{2 \rho}{\rho - \Delta + 1}\Big) \nn
&- 2\Li\Big(\frac{2 \rho}{\rho - \Delta - 1}\Big) 
- 2\Li\Big(\frac{2 \rho}{\rho + \Delta + 1}\Big) \nn
&- \log\frac{(\Delta + \rho)^2 - 1}{(\Delta - \rho)^2 - 1} 
\,\log\!\bigg(\frac{\Delta^2 - (1 + \rho)^2}{\Delta^2 - (1 - \rho)^2} \frac{M_K^2}{M_\pi^2}\bigg)\bigg]\bigg\} ~,
\label{eq:omegaKpi} 
\end{align}
where $\Sigma = (M_K^2+M_\pi^2)/s$, $\Delta = (M_K^2-M_\pi^2)/s$, and $\rho = \lambda^{1/2}/s$.
We note that also $\omega_{K\pi}$ is ultraviolet finite and does not require a counterterm.
The contribution of graphs with a virtual-photon coupling to the $K\to\pi\gamma^*$ 
vertex, see Fig.~\ref{fig:feynman2},
is essential for this cancelation to work.  This vertex, with full dependence on the 
form factor $F(s)$, can be constructed from gauge invariance requirements
along the same lines discussed in Refs.~\cite{GKPV,KMGS} (for the bremsstrahlung parts
in $K_{\ell3\gamma}$).
We find therefore that \emph{all} graphs contributing to radiative corrections in $d\Gamma/ds$
as shown in Figs.~\ref{fig:feynman1} and \ref{fig:feynman2} depend on the full form factor $F(s)$ without constraints
on its functional form, so $F(s)$ can be fully factorized, and the correction factors $\Omega$ are independent of it.
[This of course ceases to be true once ``internal'' radiative corrections are included as e.g.\ photon exchange
within a charged meson loop contribution to the form factor; see Ref.~\cite{Photons}.]

We also show the correction factor $\Omega(K^+\to\pi^+ \mu^+\mu^-)$ 
(in the soft-photon approximation) in Fig.~\ref{fig:Kmu}.
The additional term is seen to increase it compared to the neutral kaon channel by a smooth function
contributing about 1\% to 1.5\%.

\subsection{\boldmath{$K\to\pi e^+e^-$}}

Inserting the electron mass $m_e$ into the correction factor in the soft approximation,
Eq.~\eqref{eq:Ollsoft}, or the exact form in Appendix~\ref{app:Ollexact}, leads to two
observations that may seem unsettling at first sight. First, the correction factor
$\Omega_{e^+e^-}$ is large, up to 10\% or so, despite the fact that the Coulomb pole 
that is the dominant effect in the muon channel is hardly visible, as its characteristic
extension is now given by the scale of the electron mass and hence tiny.  
Second, there are much more severe deviations between the exact form and the 
soft-photon approximation, even though both lead to results larger than na\"ively expected.

The explanation for this behavior can be found by expanding $\Omega_{e^+e^-}$ from Eq.~\eqref{eq:Ollexact}
in the electron mass, neglecting terms of $\Order(m)$, which leads to
\begin{align}
\Omega_{e^+e^-} &= \frac{\alpha}{\pi} \bigg\{\frac{1}{4} - 2 \bigg[\log\epsilon 
+ \frac{(1-\epsilon)(3-\epsilon)}{4}\bigg]\log\delta  \nn
&+\frac{1-\epsilon}{2} \bigg[(3-\epsilon)\log\left(1-\epsilon\right) - 
\frac{11 - 3\epsilon}{2}\bigg] \nn
&- 2 \log\epsilon + \frac{\pi^2}{3}-2\Li\left(\epsilon\right) 
\bigg\} + \Order(m)~, \label{eq:Sudakov} 
\end{align}
with $\epsilon = 2\Em/\sqrt{s}$ and $\delta=m^2/s$.  
The result for the $\log\delta$ enhanced terms is in agreement with Ref.~\cite{Bell}.
Numerically, one finds that
this reproduces the exact form to an accuracy of $10^{-6}$.  
Equation~\eqref{eq:Sudakov} contains terms that diverge logarithmically in the limit
of the electron mass approaching zero, so-called Sudakov logarithms.  
This is not in contradiction with the Kinoshita--Lee--Nauenberg theorem~\cite{Kinoshita,LeeNauenberg},
which only states the absence of such mass singularities in \emph{inclusive} transition probabilities.
Indeed, it is easy to check that in the inclusive limit, corresponding to $\epsilon=1$ in 
Eq.~\eqref{eq:Sudakov}, all terms $\propto \log\delta$ vanish, and the whole correction factor
reduces to $\Omega_{e^+e^-} = \alpha/4\pi+\Order(m)$.  
We also understand why the soft-photon approximation is insufficient: it amounts to neglecting all terms
of $\Order(\epsilon)$ in Eq.~\eqref{eq:Sudakov}, which changes the coefficient of the 
mass logarithm and fails to yield the required cancelation in the inclusive limit.

\begin{figure}
 \centering
 \includegraphics[width=\linewidth]{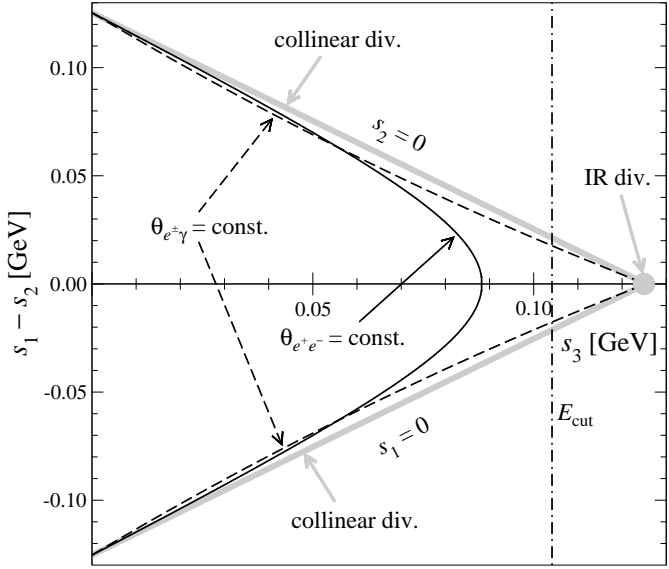}
 \caption{Dalitz plot for $\gamma^*\to e^+e^-\gamma$ at maximal $s=(M_K-M_\pi)^2$. 
  Shown are the positions of the collinear singularities for $s_1=0$ and $s_2=0$ (broad grey bands)
  as well as the infrared singularity at their intersection (thick grey dot).
  Drawn relative to those are various cut positions in $\theta_{e^\pm\gamma}$ (dashed line), $\Em$ (vertical dash-dotted), 
  and $\theta_{e^+e^-}$ (full black).}
 \label{fig:Dalitz}
\end{figure}
The origin of the electron mass singularities is well understood: they stem from 
photons radiated collinearly off the (nearly massless) electrons.  
The close-to-on-shell electron propagator in a bremsstrahlung diagram, 
with a photon of momentum $l$, is of the form
\beq
\frac{1}{2p_-l} = \frac{1}{2\left(\sqrt{m^2+\vec{p}_-^2}-|\vec{p}_-|\cos\theta_{e^-\gamma}\right)|\vec{l}|} ~,
\eeq
which in the limit $m\to0$ diverges in forward direction, $\cos\theta_{e^-\gamma} =1$.
To illuminate the structure of the various divergences, we show a sketch of the 
Dalitz plot for the decay $\gamma^*(p-k)\to e^+(p_+)e^-(p_-)\gamma(l)$ in Fig.~\ref{fig:Dalitz}, 
for a ``mass'' of the decaying virtual photon corresponding to the kinematic maximum
$\sqrt{s_{\rm max}}=M_K-M_\pi$.  We discuss this sub-decay in terms of three Mandelstam variables
$s_{1/2}=(p_\mp+l)^2$, $s_3=(p_++p_-)^2$, obeying $s_1+s_2+s_3=s+2m^2\approx s$ in the massless limit,
where the Dalitz plot degenerates into a triangle.  
Figure~\ref{fig:Dalitz} shows the collinear singularities occurring along the lines $s_1\approx 0$, $s_2\approx 0$, 
as well as the infrared singularity at $l=0$ at the intersection of those two lines.
It is obvious that excluding all photons with an energy above $\Ec$ leaves mass singularities
in the resulting corrections, stemming precisely from hard collinear photons.

\begin{sloppypar}
However, much as the inclusion of soft photons in a realistic measurement is an experimental
requirement due to finite detector energy resolution, so is the inclusion of collinear photons
due to finite angular resolution.
We therefore have to cut not only on the photon energy, but also on the angles 
$\theta_{e^\pm\gamma}$.\footnote{This of course also holds true in principle for
bremsstrahlung off heavy particles, such as the muon decay channel discussed in the previous section.
However, without any strong collinear enhancement, the excision of photons at relative
angles below, say, $20^\circ$ from the full solid angle leads to a modification
of the correction factor below the permille level---a negligible effect.}
Figure~\ref{fig:Dalitz} shows the position of such an angular cut in the $e^+e^-\gamma$ Dalitz plot:
obviously, the hard collinear photons are now part of the correction factor.
The result for $\Omega_{e^+e^-}$ with such an angular cut is given,
in the massless limit ($m=0$), as
\begin{align}
\Omega_{e^+e^-}^{e\gamma} &= \frac{\alpha}{\pi} \bigg\{\frac{1}{4} 
- 2\bigg[\log\epsilon + \frac{(1 - \epsilon) (3 - \epsilon)}{4}\bigg]\log\frac{1 - c_{e\gamma}}{1 + c_{e\gamma}} 
\nn
&- \frac{1 - \epsilon}{2} \bigg[(3 - \epsilon)\log\left(1 - \epsilon\right)  
- \frac{11 - 3 \epsilon}{2}- \frac{8}{1 - c_{e\gamma}} \nn
& \quad \times \bigg(\frac{1 + c_{e\gamma}}{(1 - \epsilon)(1 - c_{e\gamma})}
\log\Big(\epsilon + \frac{2(1 - \epsilon)}{1 + c_{e\gamma}}\Big) - 1\bigg)\bigg] \nn
& - \frac{\pi^2}{3} + 2\Li\left(\epsilon\right) \bigg\} ~, \label{eq:cuteg} 
\end{align}
with $c_{e\gamma} = \cos\theta_{e^\pm\gamma}$.  
Equation~\eqref{eq:cuteg} holds for realistic (small) angular cuts;
for $\cos\theta_{e^\pm\gamma} < -\epsilon/(2-\epsilon)$, i.e.\ basically with all photons radiated
in the forward direction relative to electron and positron cut out, 
the angular cut alone becomes more restrictive than the photon-energy cut, 
and the above formula has to be continued with the replacement $\epsilon\to-2c_{e\gamma}/(1-c_{e\gamma})$.
The result then translates smoothly into the fully inclusive one for $c_{e\gamma}\to-1$.
On the other hand, for small angles or $c_{e\gamma}\approx 1$, Eq.~\eqref{eq:cuteg} shows how the 
mass singularity in Eq.~\eqref{eq:Sudakov} is translated into a kinematical or phase space singularity.
\end{sloppypar}

\begin{figure}
 \centering
 \includegraphics[width=\linewidth]{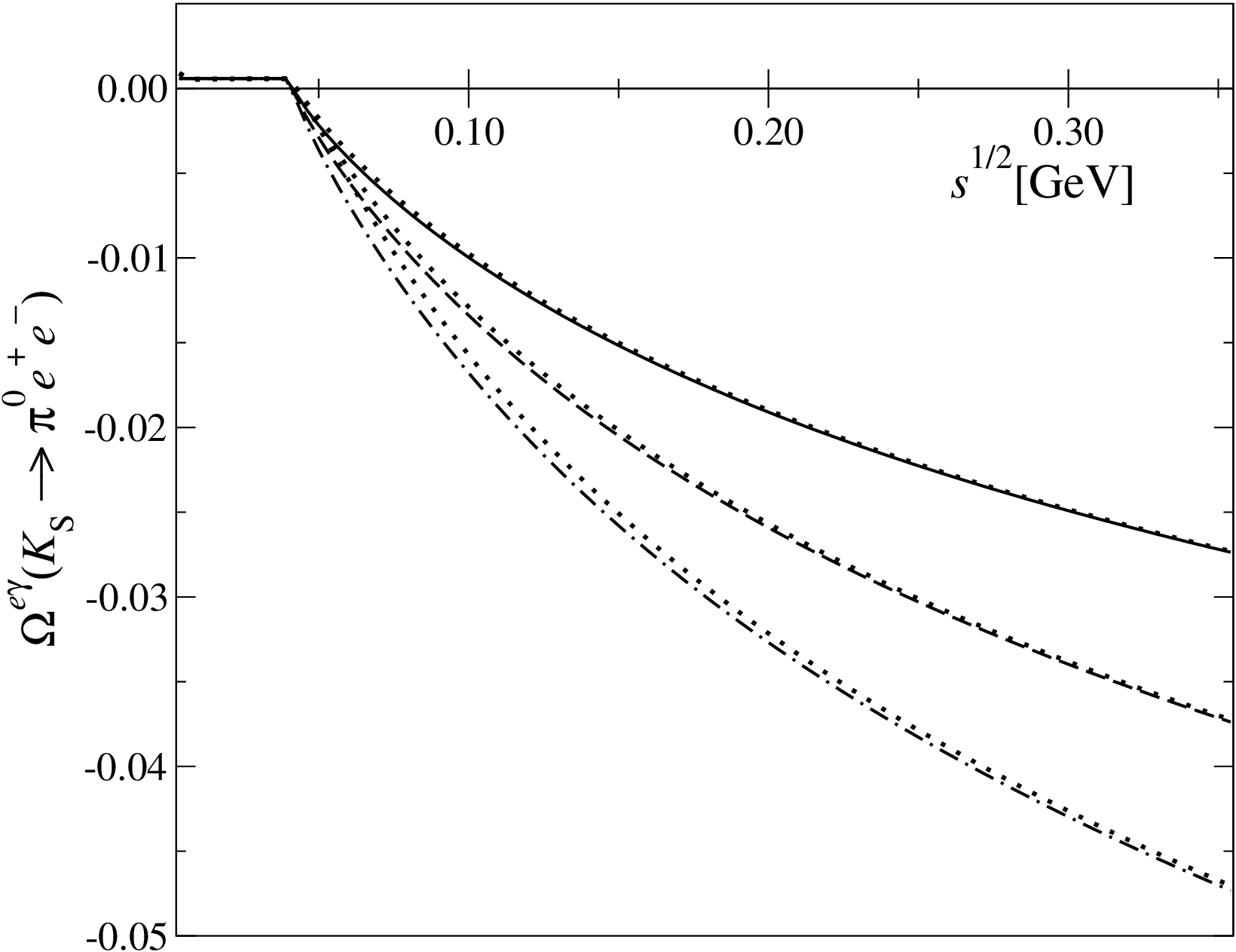}
 \caption{Correction factor for $d\Gamma/ds(K_S\to\pi^0 e^+e^-)$ in the massless limit, with cuts on 
 $\theta_{e^\pm\gamma} = 20^\circ$ (full line), $10^\circ$ (dashed), and $5^\circ$ (dash-dotted), all with $\Ec=20$~MeV.
 For comparison, the dotted lines nearby show the same curves for the physical electron mass.}
 \label{fig:egCut}
\end{figure}
$\Omega_{e^+e^-}^{e\gamma}$ is displayed graphically for three different angular cuts in Fig.~\ref{fig:egCut},
all with $\Ec=20$~MeV.  For comparison, Fig.~\ref{fig:egCut} also shows, in addition to the massless
result Eq.~\eqref{eq:cuteg}, the exact result with finite electron mass, obtained numerically:
the differences are tiny.  Obviously the correction factor is largest with the tightest cut on the angle,
however for realistic angles, it is small over the whole dilepton distribution range, on the level of
2\% to 4\%.

\begin{figure}
 \centering
 \includegraphics[width=\linewidth]{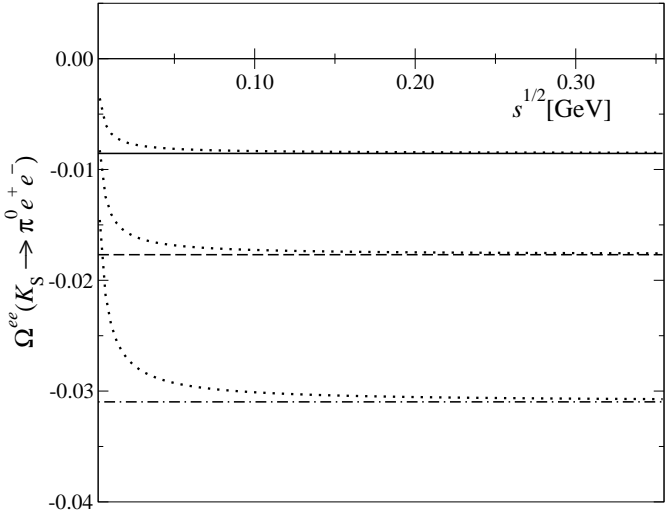}
 \caption{Correction factor for $d\Gamma/ds(K_S\to\pi^0 e^+e^-)$ in the massless limit, with cuts on 
 $\theta_{e^+e^-} = 160^\circ$ (full line), $170^\circ$ (dashed), and $175^\circ$ (dash-dotted).
 For comparison, the dotted lines nearby show the same curves for the physical electron mass.}
 \label{fig:epemCut}
\end{figure}

There is another possibility how to apply phase space cuts such as to cover all the singular
regions in the bremsstrahlung that is maybe slightly less obvious, namely to
cut on the positron--electron angle $\theta_{e^+e^-}$.  All singularities, both the infrared
divergence and the collinear ones, occur for electron and positron back-to-back, i.e.\ 
$\cos\theta_{e^+e^-}\approx -1$.  A line of constant $\theta_{e^+e^-}$ is also shown
in Fig.~\ref{fig:Dalitz}.
The result for $\Omega_{e^+e^-}$ with such an $e^+e^-$ collinear cut applied,
again in the massless limit, is given by
\begin{align}
\Omega_{e^+e^-}^{ee} &= \frac{\alpha}{\pi} \bigg\{\frac{1}{4} 
+ \Li\Big(\frac{c_{ee}-1}{c_{ee}+1}\Big) - \frac{9-5c_{ee}}{4(1-c_{ee})}\nonumber\\
&-2  \frac{2 - c_{ee}}{(1 - c_{ee})^2}\log\frac{1 + c_{ee}}{2} \bigg\} ~, \label{eq:cutee}
\end{align}
with $c_{ee} = \cos\theta_{e^+e^-}$.
Equation~\eqref{eq:cutee} again is in accordance 
with Ref.~\cite{Bell}.  With no cut on any dimensionful quantity applied (and $m=0$), 
$\Omega_{e^+e^-}^{ee}$ is bound to be a constant.  Comparison to results at finite electron
mass obtained numerically for three different cut angles, see Fig.~\ref{fig:epemCut},
show some deviations for very small $s$.  Again, with realistic angular cuts, the resulting
correction factors are small on the few-percent level.

Finally, after this extensive discussion of $\Omega_{e^+e^-}$ with various cuts and approximations, which
directly yields the correction factor $\Omega(K_S\to\pi^0e^+e^-)$, 
we have to remark on the phenomenologically most relevant case, $K^+ \to \pi^+ e^+ e^-$.
As for $K^+ \to \pi^+ \mu^+ \mu^-$, we have to add the corrections on the hadronic side $\Omega_{K\pi}=\omega_{K\pi}$
to obtain the full result for $d\Gamma/ds$.  
It is fully justified to use the soft-photon approximation Eq.~\eqref{eq:omegaKpi} for this part, 
as it does not involve the lepton mass at all and hence shows no enhancement for $m\to 0$.
An analytic implementation of angular cuts in the $e^+e^-\gamma$ system on $\Omega_{K\pi}$ is not feasible,
however it is also not necessary, as there are no collinear enhancements, and the angular cuts only
remove a very small part of the solid angle.  Estimating the effect of a cut on the lepton--photon 
angle of $20^\circ$ leads to a modification of the correction factor of a few times $10^{-4}$
and can hence safely be disregarded.

\section{\boldmath{$\ell^+\ell^-$} asymmetry}\label{sec:asym}

\begin{figure}
\centering
\includegraphics[width=0.95\linewidth]{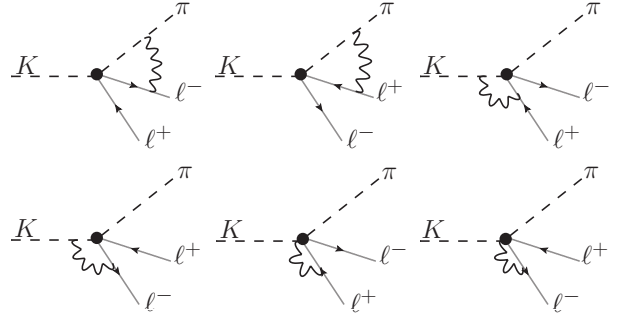}
\caption{Feynman graphs contributing to the $\ell^+\ell^-$ asymmetry in 
$K_+\to\pi^+\ell^+\ell^-$ through virtual-photon corrections.
The real-photon contributions via interference of lepton and meson bremsstrahlung, 
see Figs.~\ref{fig:feynman1} and \ref{fig:feynman2} respectively, are not displayed again.}
\label{fig:feynmanAsym}
\end{figure}
There is a set of radiative corrections in $K^+\to\pi^+\ell^+\ell^-$ not considered so far,
namely virtual-photon corrections linking the hadronic and the leptonic currents,
see Fig.~\ref{fig:feynmanAsym}, plus the interference terms of bremsstrahlung from
leptons with bremsstrahlung from mesons.  All these contributions are odd under exchange
of $\ell^+$ and $\ell^-$ and hence cancel in $d\Gamma/ds$, see Eq.~\eqref{eq:Omegadef}.  
However, they induce an $\ell^+\ell^-$ (or $t-u$) asymmetry, 
which we define in a normalized way as
\beq
A_\nu(s) = \bigg(\frac{d\Gamma}{ds}\bigg)^{-1} \int d\nu \,{\rm sgn}(\nu) \frac{d^2\Gamma}{ds\,d\nu} ~,\label{eq:Adef}
\eeq
where ${\rm sgn}(x)$ is the sign function.  
With this definition, the asymmetry is independent of overall coupling constants, and we expect it
to be on the typical level of radiative correction effects, i.e.\ a few percent at most.
Referring back to Eq.~\eqref{eq:DalitzEM}, we can re-express
$A_\nu$ in a very simple form in terms of the correction factors $\hat\omega$ and $\tilde\omega$ as
\beq
A_\nu(s) = \frac{1}{\lambda^{3/2}\sigma(1-\sigma^2/3)} 
\int_0^{\sigma\lambda^{1/2}} \!\!\!\! d\nu\big[\big(\lambda-\nu^2\big)\hat\omega + m^2\nu\tilde\omega\big]~. 
\label{eq:Aomega}
\eeq

The explicit results for $\hat\omega$ and $\tilde\omega$ are significantly more complicated and lengthy than those shown so far
and are therefore relegated to Appendix~\ref{app:asym}.  
Several features of these corrections however need to be discussed in some detail.
First, in contrast to what we found for the corrections to the dilepton spectra, $\hat\omega$ and $\tilde\omega$
are not automatically ultraviolet finite.
Naturally, nothing is known about the finite coefficients of the required counterterms, 
whose effects we therefore have to estimate by different means.  We do this by employing the corresponding
ultraviolet scale dependence of the loop contributions: choosing the mass of the $\rho$ as a natural central scale,
we vary it by a factor of $e=2.71\ldots$, i.e.\ between $0.37M_\rho$ and $2.71M_\rho$.  We believe this 
to be a reasonable uncertainty estimate.

Second, also in contrast to the correction factors for the dilepton distributions, 
the dependence on the form factor $F(s)$ does not factorize any longer in the loop contributions
of Fig.~\ref{fig:feynmanAsym}.  In fact, a hadronic loop in $F(s)$ would immediately lead to non-trivial
(i.e.\ non-factorizing) two-loop graphs for the photon corrections.  
As already indicated in the introduction, we approximate such effects by assuming a linear form factor,
with the slope as measured in Ref.~\cite{Batley:2009pv}, $F'(0)/F(0) = (2.32\pm0.18)M_K^{-2}$.  
The correction factors $\hat\omega$, $\tilde\omega$ then contain terms that depend on $F'(s)/F(s)$.

There is another subtlety in the calculation of the graphs in Fig.~\ref{fig:feynmanAsym}.
We have always drawn Feynman diagrams with point-like $K\pi\ell^+\ell^-$ vertices, owing to the fact
that the first term in the form factor definition Eq.~\eqref{eq:FF} contains a factor of $s$, 
canceling the photon propagator.  In contrast to all graphs in Figs.~\ref{fig:feynman1} and \ref{fig:feynman2},
the second term in the form factor $\propto (k^2-p^2)(k-p)^\mu$, which does not contain such a factor of $s$,
does not vanish any more diagram by diagram for loop contributions with photons linking mesons to fermions.
The cancelation of apparent two-photon-cut contributions in the full result 
is hence a useful check on the calculation.

Finally, we only include bremsstrahlung in the soft-photon approximation (thereby in particular eschewing 
the non-trivial question of how to properly \emph{define} the asymmetry, once the real-photon momentum
is not neglected in the kinematic relations), also for the electron--positron final state.  
We deem this justified for two reasons: first, we find that the electron mass or collinear singularities
actually cancel in the asymmetry Eq.~\eqref{eq:Aomega}, hence there is a smooth limit of the electron mass
approaching zero, no unnatural enhancement occurs, and there is no need to further cut on the angles
in the phase space integral.  Second, the remaining uncertainty inherent in the soft approximation
we consider well-covered by the sizeable (as we shall see below) error due to the ultraviolet scale dependence
discussed above. 

\begin{figure}
\centering
\includegraphics[width=\linewidth]{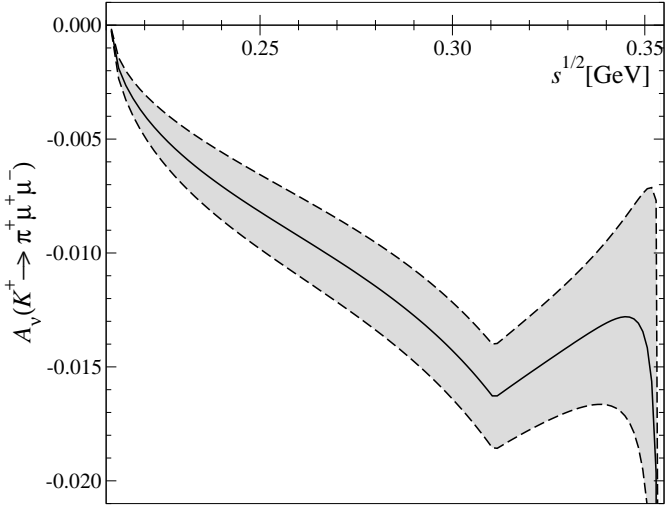}
\caption{The asymmetry $A_\nu(K^+\to\pi^+\mu^+\mu^-)$ according to Eq.~\eqref{eq:Adef}.
The width of the grey band mirrors the uncertainty due to an unknown counterterm,
mimicked by varying the ultraviolet scale between $0.37M_\rho$ and $2.71M_\rho$.}
\label{fig:AsymMu}
\end{figure}

\begin{figure}
\centering
\includegraphics[width=\linewidth]{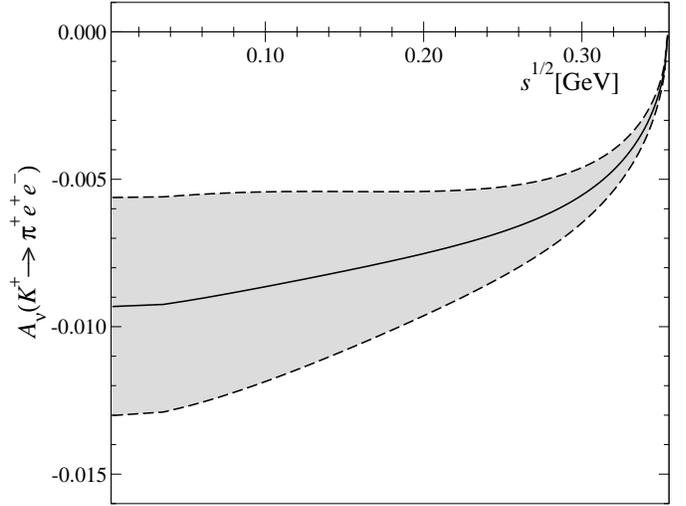}
\caption{The asymmetry $A_\nu(K^+\to\pi^+e^+e^-)$ according to Eq.~\eqref{eq:Adef}.
The width of the grey band mirrors the uncertainty due to an unknown counterterm,
mimicked by varying the ultraviolet scale between $0.37M_\rho$ and $2.71M_\rho$.}
\label{fig:AsymE}
\end{figure}

\begin{sloppypar}
The resulting asymmetries $A_\nu(s)$ are displayed in Fig.~\ref{fig:AsymMu} for $K^+\to\pi^+\mu^+\mu^-$
and in Fig.~\ref{fig:AsymE} for $K^+\to\pi^+e^+e^-$.  
Both are shown for a photon-energy cut of $\Ec=20$~MeV; the variation of the cutoff 
by 10~MeV leads to changes in the asymmetries significantly below the ranges displayed
by the grey bands, which represent the estimated uncertainties due to the ultraviolet counterterm.
Similarly, the error due to the experimental uncertainty in the slope of the form factor
quoted above is totally negligible in comparison.
The main features of the results shown in Figs.~\ref{fig:AsymMu} and \ref{fig:AsymE} 
can be readily understood:
\begin{enumerate}
\item The sign of the asymmetry is negative: positive $\nu$ corresponds to large $t$ and 
small $u$, or a small invariant mass of $\pi^+$ and $\ell^+$, for which the repulsion of
the two positively charged particles in the final state dominates; vice versa, 
for negative $\nu$ or small $t$, the attraction between $\pi^+$ and $\ell^-$ leads to 
an enhancement. 
\item The asymmetry for the muon channel vanishes for small $s \to 4m^2$:
due to the integration limit $\nu_{\rm max}= \sigma\lambda^{1/2}$,
the fact that $\hat\omega$ is odd in $\nu$, plus the explicit factor of $\nu$ in front of $\tilde\omega$, 
it is obvious that the numerator in Eq.~\eqref{eq:Aomega} has one more power of $\sigma$ than the 
denominator, hence $A_\nu(s)$ vanishes square-root-like at the lower end of the spectrum.
This effect is invisible in the electron asymmetry due to the smallness of the electron mass, 
for which effectively $\sigma\approx 1$.
\item A similar argument explains the behavior at the upper end of the spectrum, for large
$s\to(M_K-M_\pi)^2$: were it only for the correction factor $\hat\omega$, the numerator 
would be suppressed compared to the denominator by a factor of $\lambda^{1/2}$, and the asymmetry
would approach zero; however the term $\propto\tilde\omega$, when integrated over $\nu$, is \emph{enhanced} 
by a factor of $\lambda^{1/2}$, hence the divergence seen in the muon asymmetry.
On the other hand, $\tilde\omega$ comes with a prefactor of $m^2$, such that this term is negligible
in the electron channel, where we therefore indeed see the asymmetry going to zero at the upper limit of phase space.
\item Finally, there appears to be a cusp-like structure in the muon asymmetry, Fig.~\ref{fig:AsymMu}.
This can be traced back to the $\nu$ integration hitting the Coulomb poles in the $t$ and $u$ channels,
which happens for $s^*=m(\mk-\mpi)/(M_\pi+m)$.  The Coulomb pole is an integrable singularity of the form
$(t-t_{\rm thr})^{-1/2}$ (at this order in $\alpha$), which is touched tangentially by the integration
(phase space) limit, leading to a non-regular behavior of the type
\beq
\int_{t_{\rm min}(s)} \frac{dt}{\sqrt{t-t_{\rm thr}}} ~\simeq~ \int_{k(s-s^*)^2} \frac{dt}{\sqrt{t}} ~\propto \sqrt{(s-s^*)^2} ~,
\eeq
hence like an absolute-value function.
Again, as $s^*$ scales with $m$, this non-regular behavior practically coincides with the threshold
in the electron channel and is therefore invisible in Fig.~\ref{fig:AsymE}.
\end{enumerate}
The overall size of the two asymmetries is indeed typical for an electromagnetic effect
on hadronic observables, it lies in the range of 1\%--2\%.  The asymmetry for the electron
channel shows no enhancement, and is virtually identical to the one in the zero mass limit.
\end{sloppypar}

\section{\boldmath{$P\to V\ell^+\ell^-$} and \boldmath{$V\to P\ell^+\ell^-$} decays}\label{sec:other}

It is obvious that any decay $P\to P'\ell^+\ell^-$ dominated by one-photon exchange,
where $P$, $P'$ denote pseudoscalar mesons, will have the 
same universal electromagnetic correction effects as discussed in the previous sections.  
They only depend on the structure of the hadronic current, Eq.~\eqref{eq:FF},
which in turn is dictated by parity and gauge invariance, so the central results
can be readily adapted for e.g.\ charm or bottom decays, with the appropriate 
mass replacements.

The transfer of our calculations to decays of the type
$P\to V \ell^+ \ell^-$ or $V\to P \ell^+ \ell^-$, for pseudoscalar $P$ and vector $V$ mesons, 
may be less obvious.  Physical processes of these types, as already mentioned in 
the introduction, include the $\pi^0$ and $\eta$ Dalitz decays 
$\pi^0\to\gamma e^+e^-$ (see e.g.\ Ref.~\cite{Kampf:2005tz} for theoretical work on this decay),
$\eta\to\gamma\ell^+\ell^-$~\cite{Bijnens:1988kx}, or the vector meson conversion decays such as 
$\omega\to\pi^0\ell^+\ell^-$ or $\phi\to\eta\ell^+\ell^-$~\cite{Terschluesen:2010ik}.
Reference~\cite{Kampf:2005tz} gives a detailed account on radiative corrections in $\pi^0\to\gamma e^+e^-$ far
beyond the universal effects considered here, however relying mainly on the soft approximation
for the real-photon radiation and without an analysis of collinear singularities and angular 
phase space cuts.

We generically denote decays of these types by $A(k)\to B(p)\ell^+(p_+)\ell^-(p_-)$, with 
$M_{A/B}$ the masses of $A$ and $B$ ($M_B=0$ for the $\pi^0$ or $\eta$ Dalitz decays).  
What is common to these channels is that they involve anomalous decay vertices
of odd intrinsic parity, so the hadronic vector current is written in terms of the Levi-Civita tensor,
\beq
\langle B(p) | J_{\rm em}^\mu(0) | A(k) \rangle ~\propto~
\epsilon^{\mu\nu\alpha\beta} \epsilon^*_\nu k_\alpha p_\beta \,F(s) ~, \label{eq:PVhadronic}
\eeq
where the form factor $F$ is again a function of $s=(k-p)^2$, and
we have spelled out the case of the $P\to V$ decay, with $\epsilon^*_\nu$ the polarization 
vector of the outgoing $B$.
Equation~\eqref{eq:PVhadronic} can be transformed 
into the $V\to P$ case by the simple replacement $\epsilon^*_\nu \to \epsilon_\nu$, $\epsilon_\nu$
then being the polarization vector of the incoming $A$.  

It is conventional to normalize the rates by the real-photon width $\Gamma_0 \doteq \Gamma(A\to B\gamma)$
(which, in contrast to $K\to\pi\gamma$, is not required to vanish by gauge invariance).
Including radiative correction factors, we can write the Dalitz plot distribution as
\begin{align}
\frac{d\Gamma}{\Gamma_0} = \frac{n\,\alpha |\bar F(s)|^2}{8\pi s(M_A^2-M_B^2)^3}\, 
&\Big[ \left(\left(2-\sigma^2\right)\lambda + \nu^2 \right)(1+\omega) \nn 
& + \left(\sigma^2\lambda - \nu^2 \right) \bar\omega \Big]
ds\,d\nu ~,\label{eq:DalitzPV}
\end{align}
where $\lambda \doteq \lambda(M_A^2,s,M_B^2)$, $\nu=t-u$ defined as before,
$\bar F(s)=F(s)/F(0)$ is the normalized form factor, and
$n=1$ for vector meson conversion and $n=2$ for pseudoscalar Dalitz decays.
$\omega$ and $\bar\omega$ are the \emph{same} corrections factors of $\Order(\alpha)$ 
as discussed in various forms and approximations in Sect.~\ref{sec:dGds}.
We only consider processes with uncharged bosons (as in all the sample processes mentioned above), 
so we do not include the correction terms odd in $t-u$.
The resulting dilepton invariant mass spectrum is 
\beq
\frac{d\Gamma}{ds\,\Gamma_0} = \frac{n\,\alpha |F(s)|^2}{2\pi s(M_A^2-M_B^2)^3}\, 
\lambda^{3/2}\sigma\Big(1-\frac{\sigma^2}{3}\Big) (1+\Omega) ~,
\eeq
where $\Omega$ is again given in terms of $\omega$ and $\bar\omega$ by the same relation 
Eq.~\eqref{eq:Omegadef} as in $K\to\pi\ell^+\ell^-$.  
So while the Dalitz plot distribution Eq.~\eqref{eq:DalitzPV} is entirely different 
from that of Eq.~\eqref{eq:Dalitz}, owing to the different spin of one of the bosons involved,
the dilepton mass spectrum and its radiative correction factor look, apart from 
the different form factors involved, the same.
We emphasize that this also holds for the bremsstrahlung contributions beyond the 
soft-photon approximation, for which this is far from obvious in the course of the calculation;
see Appendix~\ref{app:brems} for details.

\section{Summary and conclusion}\label{sec:summary}

\begin{sloppypar}
We have calculated the universal radiative corrections for meson decay processes with 
a lepton--antilepton pair in the final state, in particular the flavor-changing
neutral current process $K\to\pi\ell^+\ell^-$, with the following main results.
\begin{enumerate}
\item 
For the most interesting observable, the differential spectrum with respect to the 
invariant mass of the dilepton pair, radiative corrections at $\Order(\alpha)$ 
are given as a simple sum of corrections to the hadronic and the leptonic current,
both of which are ultraviolet finite without requiring a counterterm.
\item
For the muon final state, the soft-photon approximation with a cut on the additional
photon energy in the bremsstrahlung contribution is very accurate and leads 
to an enhancement on the 1\%--2\% level, apart from the Coulomb pole close to threshold.
\item
The (unphysical) excision of hard collinear bremsstrahlung photons leads to the survival of
logarithmic mass singularities in the electron channels.  This requires the introduction
of additional angular cuts in the leptonic radiative corrections, 
either between bremsstrahlung photon and electron/position, or on the electron--positron angle.
We provide closed analytic results for the cut dependence of these correction factors
in the limit of the electron mass going to zero, which is shown to be a very good approximation.
\item
Radiative corrections in the charged decays $K^+\to\pi^+\ell^+\ell^-$ that link
the hadronic and the leptonic current are odd under exchange of $\ell^+$ and $\ell^-$.
They cancel in the dilepton spectrum, but induce an $\ell^+\ell^-$ asymmetry 
on the level of 1\%--2\%.  This asymmetry is free of small mass singularities
and not enhanced in the electron case, but requires ultraviolet renormalization 
via a counterterm that induces a sizeable uncertainty.
\item
Finally, we have shown that the universal correction factors for the dilepton spectra can be 
directly used also for other decays such as $\pi^0$ or $\eta$ Dalitz decays 
or vector meson conversion decays like $\omega\to\pi^0\ell^+\ell^-$, despite a different
form of the hadronic vector current involved.
\end{enumerate}
These results ought to provide a useful and necessary ingredient in future precision
measurements of the decay processes concerned.
\end{sloppypar}


\begin{acknowledgement}
\begin{sloppypar}
\textbf{Acknowledgements\ }
We would like to thank Gino Isidori for suggesting to undertake this study, and for
extensive advice and helpful discussions along the way.
Furthermore, we thank Guido Bell for useful e-mail communications.
Partial financial support by the Helmholtz Association through funds provided
to the virtual institute ``Spin and strong QCD'' (VH-VI-231), 
by the Integrating Activity project ``Study of Strongly Interacting Matter''
(acronym HadronPhysics2, grant agreement No.~227431) under the Seventh 
Framework Programme of the EU,
and by DFG (SFB/TR 16, ``Subnuclear Structure of Matter'') is gratefully
acknowledged. 
\end{sloppypar}
\end{acknowledgement}


\begin{appendix}
\renewcommand{\theequation}{\thesection.\arabic{equation}}

\setcounter{equation}{0}
\section{Correction factor \boldmath{$\Omega(K_S\to\pi^0\ell^+\ell^-)$}}\label{app:Ollexact}

In this appendix we show the analytic form of the correction factor $\Omega_{\ell^+\ell^-}$ beyond the 
soft-photon approximation discussed in Sect.~\ref{sec:musoft}, i.e.\ with the exact
$\Em$ and full mass dependence, but without any additional angular cuts.  We find
{\allowdisplaybreaks
\begin{align}
\Omega_{\ell^+\ell^-} &=\Omega_{\ell^+\ell^-}^{\rm soft}
+ \frac{\alpha}{\pi} \bigg\{\frac{1+\sigma^2}{\sigma} \bigg[2 \Li\Big(\frac{1-\sigma}{
        1+\sigma}\Big) - \Li\Big(\frac{1-\sigma_\epsilon}{1+\sigma}\Big) 
   \nonumber\\
&- \Li\Big(\frac{1-\sigma}{1+\sigma_\epsilon}\Big)+ 
      \Li\Big(\frac{1+\sigma}{2}\Big) - \Li\Big(\frac{1+\sigma_\epsilon}{2}\Big)  
       \nonumber\\
&- \Li\Big(\frac{1-\sigma}{2}\Big)
+ \Li\Big(\frac{1-\sigma_\epsilon}{2}\Big) 
      +\Li\Big(\frac{1+\sigma_\epsilon}{1+\sigma}\Big) 
       - \frac{\pi^2}{3}\nonumber\\& 
      +\Li\Big(\frac{1-\sigma}{1-\sigma_\epsilon}\Big) 
      -\log\left(\frac{1+\sigma}{1-\sigma} \frac{1-\sigma_\epsilon}{1+\sigma_\epsilon}\right) 
      \log\frac{2 \epsilon}{1+\sigma}\bigg] 
    \nonumber\\
&+ 4 \log\frac{\sigma + \sigma_\epsilon}{2 \sigma} 
- \frac{\epsilon}{\sigma} \left(2 - \frac{\epsilon}{3 - \sigma^2}\right) 
    \log\frac{1+\sigma_\epsilon}{1-\sigma_\epsilon} \nn
&  + 2 \log\left(1 - \epsilon\right) 
+ \frac{9 - 2\sigma^2 + \sigma^4}{8 \sigma \left(3 - \sigma^2\right)} 
\log\left(\frac{1+\sigma}{1-\sigma} \frac{1-\sigma_\epsilon}{1+\sigma_\epsilon}\right) \nn
& + \frac{\sigma - \sigma_\epsilon}{\sigma \left(1 - \sigma_\epsilon^2\right)} 
\bigg[  \left(1+\sigma \sigma_\epsilon\right) \left(4 + \frac{3 \left(1 - \epsilon\right)}{
         2 \left(3 - \sigma^2\right)}\right) \nn
& - \frac{\left(9 + \sigma^2\right) \sigma \sigma_\epsilon}{4 \left(3 - \sigma^2\right)} 
- \frac{5}{4}  \bigg]\bigg\} ~,\label{eq:Ollexact}
\end{align}}\noindent
with $\sigma_\epsilon = \sqrt{(\sigma^2-\epsilon)/(1-\epsilon)}$ and
$\epsilon = 2\Em/\sqrt{s}$.
It is easy to see from Eq.~\eqref{eq:Ollexact} that 
$\Omega_{\ell^+\ell^-} -\Omega_{\ell^+\ell^-}^{\rm soft} = \Order(\Em)$.
The numerical evaluation and comparison of this result to the soft approximation for the muon final state
is shown in Fig.~\ref{fig:Kmu}. 
Equation~\eqref{eq:Ollexact} also allows to easily derive e.g.\ the fully inclusive correction factor,
i.e.\ without phase space cuts on the real-photon emission, for finite mass $m$;
this corresponds to the limit $\epsilon=\sigma^2$.

\setcounter{equation}{0}
\section{Contributions to the \boldmath{$\ell^+\ell^-$} asymmetry}\label{app:asym}

\begin{sloppypar}
The radiative correction factors $\hat\omega$ and $\tilde\omega$, defined in Eq.~\eqref{eq:DalitzEM}
and discussed numerically in Sect.~\ref{sec:asym}, 
are given analytically by the expressions
 \begin{align}
\hat\omega &= 8\pi\alpha
\bigg\{ H^{(1)}_\pi(u)-H^{(1)}_\pi(t)+H^{(1)}_K(u)-H^{(1)}_K(t) \nn
&+ \frac{F'(s)}{F(s)}
   \left[V^{(1)}_{\pi K}(u)-V^{(1)}_{\pi K}(t)+V^{(1)}_{K\pi}(u)-V^{(1)}_{K\pi}(t)\right] \bigg\} \nn
&+\frac{\alpha}{\pi}\int_{-1}^1d\xi \Big[F_\pi(t,\xi)-F_\pi(u,\xi)
+F_K(u,\xi)-F_K(t,\xi)\Big] ~,\nn
\tilde\omega
&=  64\pi\alpha \bigg\{
 H^{(2)}_\pi(t)+H^{(2)}_\pi(u)+H^{(2)}_K(t)+H^{(2)}_K(u) + \frac{3\Delta_\ell}{m^2} \nn
& - \frac{1}{16\pi^2}  
+\frac{F'(s)}{F(s)} \bigg[
   V^{(2)}_{\pi K}(t)+V^{(2)}_{\pi K}(u) +V^{(2)}_{K\pi}(t) +V^{(2)}_{K\pi}(u)  \nn
& +4\Delta_\ell+\frac{m^2}{8\pi^2}-s\bigg(\frac{5\Delta_\ell}{3m^2}+\frac{1}{18\pi^2}\bigg)\bigg] \bigg\} ~.
\label{eq:omegaAsym}
\end{align}
As discussed in the main text, the $K\to\pi\gamma^*$ form factor is approximated 
by a linear form, inducing the terms $\propto F'(s)/F(s)$ in Eq.~\eqref{eq:omegaAsym}.
We have defined the following combinations of loop functions: 
{\allowdisplaybreaks
\begin{align}
H^{(1)}_a(x) &= (x-m^2-M_a^2) 
\left[2G_a(x)-G^+_a(x)+3G^-_a(x)\right] \nn
-2\big(&M_a^2+m^2\big)G^+_a(x)-2\big(M_a^2-m^2\big)G^-_a(x)-\bar J_a(x) ~, \nn
H^{(2)}_a(x) &= (x-m^2)\left[G^+_a(x)-G^-_a(x)\right] ~, \nn
V^{(1)}_{ab}(x) &=  
\left(M_b^2-M_a^2\right) \left(m^2+M_a^2+x\right) G_a^{(1)}(x) \nn
&+\frac{1}{4}\left(M_b^2-M_a^2-2s\right)\left(7m^2+3M_a^2-5x\right)G^-_a(x) \nn
&+\frac{1}{2}\Big[s\left(x+m^2+M_a^2\right) \nn
&\quad-\left(M_b^2-M_a^2\right)\left(3(x-m^2)-M_a^2\right)\Big] G^+_a(x) \nn
& + \frac{1}{2}\bigg[\frac{(M_b^2-M_a^2-2m^2)(M_a^2-m^2)}{x} + s+2x \nn
& \quad +M_b^2-3M_a^2-4m^2 \bigg] \bar J_a(x) -x \,\frac{\Delta_a-\Delta_\ell}{M_a^2-m^2} ~, \nn
V^{(2)}_{ab}(x) &= (x-m^2)\bigg[\left(M_a^2-M_b^2\right)G_a^{(1)}(x) \nn
&+\frac{1}{4}\left(M_a^2-M_b^2+2s\right)G_a^-(x)-\frac{s}{2}G_a^+(x) \nn
&+\frac{x+M_a^2-m^2}{2x}\bar J_a(x) -\frac{\Delta_a-\Delta_\ell}{2(M_a^2-m^2)}\bigg] ~. 
\end{align}}\noindent
Here, the tadpole loop function contains the ultraviolet divergence in dimensional regularization,
\begin{align}
  \Delta_a &= 2M_a^2\left[L+\frac{1} {16\pi^2}\log\frac{M_a}{\mu}\right] ~, \nn
  L &= \frac{\mu^{d-4}}{16\pi^2}\left[\frac{1}{d-4}+\frac{1}{2}(\gamma_E-1-\log4\pi)\right] ~,
 \end{align}
$\mu$ is the running (ultraviolet) scale, and we have made implicit use
of the notation $M_\ell \doteq m$ for the lepton tadpole function $\Delta_\ell$ above.
Of the loop functions $\bar J_a(s)$ and $G_a(s)$, 
we only need the real parts
in the two kinematical regimes $0<s\leq(M_a-m)^2$ or $s\geq(M_a+m)^2$,
where they are given by 
\begin{align}
{\rm Re} \bar J_a(s) &= \frac{1}{16\pi^2}\bigg\{
  1+\left[\frac{M_a^2-m^2}{s}-\frac{M_a^2+m^2}{M_a^2-m^2}\right]\log\frac{m}{M_a} \nonumber\\
  &-\frac{\sqrt{\lambda}}{2s}\log\frac{s-M_a^2-m^2+\sqrt{\lambda}}{s-M_a^2-m^2-\sqrt{\lambda}}\bigg\} ~,
 \end{align}
with the K\"all\'en function $\lambda = \lambda(s,M_a^2,m^2)$ used here and below,
and by
\begin{align}
{\rm Re}\, G_a(s) &= 
\frac{1}{32\pi^2\lambda^{1/2}}\bigg\{2\bigg(\log\frac{1-z_2}{1-z_1}+\log\frac{z_1}{z_2}\bigg)
\log\frac{2\Em}{\sqrt{s}} \nn
&+\Li\Big(\frac{z_2-z_1}{z_2}\Big) -\Li\Big(\frac{z_1-z_2}{z_1}\Big) +\Li\Big(\frac{z_2-z_1}{1-z_1}\Big)\nn
&-\Li\Big(\frac{z_1-z_2}{1-z_2}\Big) 
+\log^2|1-z_1|-\log^2|1-z_2| \nn
&+\log^2|z_2| -\log^2|z_1| -2\pi^2 \, \theta\big(s-(M_a+m)^2\big) \!\bigg\} \, ,
 \end{align}
respectively,
where $\theta(x)$ is the Heaviside function, and
\beq
  z_{1/2} = \frac{1}{2s}\Big(s+M_a^2-m^2 \mp \sqrt{\lambda} \Big) ~.
\eeq
Note that we already have absorbed the infrared-divergent part of the bremsstrahlung
in $G_a(s)$ in order to obtain a finite, regulator-independent function.
The remaining loop functions used above are given by
{\allowdisplaybreaks
\begin{align}
  G_a^+(s) &= \frac{1}{\lambda}\bigg[\big(m^2-M_a^2\big)\Big(\bar J_a(s)-\frac{1}{16\pi^2}\Big)
  \nonumber\\&+\frac{s-M_a^2-m^2}{16\pi^2}\log\frac{M_a}{m}\bigg] ~,\nn
  G_a^-(s) &= \frac{1}{\lambda}\bigg[s\Big(\bar J_a(s)-\frac{1}{16\pi^2}\Big) \nn
&  -\frac{s(M_a^2+m^2)-(M_a^2-m^2)^2}{16\pi^2(M_a^2-m^2)} \log\frac{M_a}{m} \bigg] ~, \nn
  G_a^{(1)}(s) &= \frac{1}{4s}\bigg\{\big(M_a^2-m^2\big)G_a^+(s) - \bar J_a(s) \nn
& +\frac{1}{16\pi^2}\bigg( \frac{M_a^2+m^2}{M_a^2-m^2}\log\frac{M_a}{m}-1\bigg) \bigg\} ~.
\end{align}}\noindent
Finally, the (finite) bremsstrahlung contribution given (in the soft-photon approximation) in terms of the 
one-parameter integral over $\xi$ in Eq.~\eqref{eq:omegaAsym} is specified by
\begin{align}
  F_a(x,\xi) &= -\frac{x-M_a^2-m^2}{p_a^2} \frac{d_a}{n_a}\log\frac{d_a+n_a}{d_a-n_a}~, \nn
  p_{\pi/K}^2 &= (1+\xi)^2M_{\pi/K}^2+(1-\xi)^2m^2 \nn & \pm(1-\xi^2)\left(x-M_{\pi/K}^2-m^2\right) ~, \nn
  d_{\pi/K} &= (1+\xi)\left(M_K^2-M_\pi^2\mp s\right)+(1-\xi) s ~,\nn
  n_{\pi/K} &= \sqrt{d_{\pi/K}^2 - 4s\, p_{\pi/K}^2 } ~.
 \end{align}
As in the main text, these finite bremsstrahlung contributions are understood
to be calculated for the photon-energy cutoff applied in the $\ell^+\ell^-\gamma$
rest frame, or $\vec{k}-\vec{p}=0$.

\setcounter{equation}{0}
\section{Details on bremsstrahlung}\label{app:brems}

In this appendix, we show some details on the calculation of bremsstrahlung
contributions to $\Omega_{\ell^+\ell^-}$ beyond the soft-photon approximation.
We concentrate on the results for the correction factor to $d\Gamma/ds$.
To describe the decay $K(k) \to \pi(p) \ell^+(p_+)\ell^-(p_-)\gamma(l)$, we make use of the extended
set of variables
\begin{align}
  s &= (k-p)^2 ~,     & s_1 &= (p_-+l)^2 ~, & s_2 &= (p_++l)^2 ~, \nn
s_3 &= (p_++p_-)^2 ~, & t_1 &= (p+p_+)^2 ~, & t_2 &= (k-p_-)^2 ~,\nn
u_1 &= (k-p_+)^2 ~, & u_2 &= (p+p_-)^2 ~, 
\end{align}
which are subject to the constraints
\begin{align}
s_1 + s_2 + s_3 &= s+2m^2 ~, \nn
s+t_i + u_i &= \mk + \mpi + m^2 + s_i ~, ~~ i=1,2 ~.
\end{align}
We also use the combinations of variables 
\beq
\nu_i = t_i-u_i ~,~~ i=1,2~, ~~ \Sigma = \nu_1+\nu_2 ~, ~~ \Delta = \nu_1-\nu_2 ~,
\eeq
and, as in the main text, $\lambda \doteq \lambda(\mk,s,\mpi)$.
The contribution resulting from the bremsstrahlung diagrams in Fig.~\ref{fig:feynman1}
can be formulated as 
\begin{align}
\Omega_{\ell^+\ell^-}^{\rm brems} &= \frac{3\alpha}{8\pi \lambda\sigma(3-\sigma^2)s} \int \frac{d\Omega}{4\pi} ds_1 ds_2
\bigg\{ \frac{N_1}{(s_1-m^2)^2}  \nn 
&  + \frac{N_2}{(s_2-m^2)^2} + \frac{N_3}{(s_1-m^2)(s_2-m^2)} \bigg\} ~. \label{eq:bremsOmega}
\end{align}
Here, $s_1$ and $s_2$ are the chosen invariant variables for the integration of the 
Dalitz plot in the $\ell^+\ell^-\gamma$ system (in the rest frame of which the calculation is performed), 
and $d\Omega$ is the solid angle integral
of this system relative to the vector $\vec{k}=\vec{p}$.  
The numerators $N_i$ in Eq.~\eqref{eq:bremsOmega}, $i=1,2,3$, are obtained by performing 
the spin and polarization sums, 
\begin{align}
N_1 &= -4\Big\{ 2m^2\big(\lambda-\nu_1^2\big) + \left(s_1-m^2\right)\left(\mk-\mpi+\nu_1\right)\Delta \nn
& + \left(s_1-m^2\right)\left(s_2-m^2\right)\left[2\left(\mk+\mpi\right)-s\right] \Big\} ~,\nn
N_2 &= -4\Big\{ 2m^2\big(\lambda-\nu_2^2\big) + \left(s_2-m^2\right)\left(\mk-\mpi-\nu_2\right)\Delta \nn
& + \left(s_1-m^2\right)\left(s_2-m^2\right)\left[2\left(\mk+\mpi\right)-s\right] \Big\} ~,\nn
N_3 &= 4\bigg\{ 2\left(s_3-2m^2\right)\left(\lambda-\nu_1\nu_2\right) - s_3 \Delta^2 + \frac{s_3-s}{2}\Sigma^2 \nn
& - \Big(\mk-\mpi+\frac{\Delta}{2}\Big)(s_1-s_2)\Sigma \bigg\} ~. \label{eq:bremsTraces}
\end{align}
Note that the non-trivial kinematic prefactors in Eq.~\eqref{eq:bremsOmega} only stem from the normalization
to the non-radiative decay spectrum; once this is reversed, Eqs.~\eqref{eq:bremsOmega} and \eqref{eq:bremsTraces}
contain the complete information to reconstruct the full kinematic dependence of the decay distribution of the bremsstrahlung
process.

The variables with non-trivial angular dependence are $\nu_{1/2}$ or $\Sigma$, $\Delta$, 
respectively, which upon integration over the solid angle yield
\begin{align}
\int&\frac{d\Omega}{4\pi} \Big\{ \Delta,\, \Sigma,\, \Delta^2,\, \Sigma^2,\, \Delta\Sigma \Big\}  = 
\bigg\{ \left(\mk-\mpi\right)\frac{s_3-s}{s} \,, \nn
& \left(\mk-\mpi\right)\frac{s_2-s_1}{s} \,, ~
\frac{(s-s_3)^2}{s^2} \Big[\left(\mk-\mpi\right)^2+\frac{\lambda}{3}\Big] \,, \nn
&\frac{(s_1-s_2)^2}{s^2} \Big[\left(\mk-\mpi\right)^2+\frac{\lambda}{3}\Big] 
+\frac{4\lambda}{3s}\left(s_3-4m^2\right) \,, \nn
&\frac{s_1-s_2}{s^2}\Big[\left(\mk-\mpi\right)^2(s-s_3)-\frac{\lambda}{3}(s+s_3) \Big] \bigg\} ~. \label{eq:angint}
\end{align}
Putting pieces together, we find
\beq
\Omega_{\ell^+\ell^-}^{\rm brems} = \frac{\alpha}{\pi\,\sigma} \bigg\{ 
\frac{s-2m^2}{s}I_2 - 2I_1 - 2I_3 + \frac{s \,I_4}{s+2m^2} \bigg\} ~, \label{eq:bremsIntComb}
\eeq
where
\begin{align}
I_1 &= \frac{m^2}{s}\int\frac{ds_1ds_2}{(s_1-m^2)^2} ~, & I_2 &= \int\frac{ds_1ds_2}{(s_1-m^2)(s_2-m^2)} ~,\nn
I_3 &= \frac{1}{s}\int\frac{ds_1ds_2}{s_1-m^2} ~, & 
I_4 &= \frac{1}{s^2} \int ds_1ds_2 \frac{s_2-m^2}{s_1-m^2} ~.\label{eq:defIi}
\end{align}
Note that the integrals $I_{1-4}$ are symmetric under the exchange $s_1 \leftrightarrow s_2$.
We document the individual results for $I_{1-4}$ for the most interesting 
cuts for $K\to\pi e^+e^-\gamma$, in (or close to) the massless limit, 
corresponding to the full corrections factors in Eqs.~\eqref{eq:Sudakov}, \eqref{eq:cuteg}, 
and \eqref{eq:cutee}.  In all cases, we show the ``hard'' parts of the integrals, i.e.\ 
those that, in the combination Eq.~\eqref{eq:bremsIntComb}, have to be \emph{subtracted}
from the fully inclusive correction factor  $\Omega_{e^+e^-} = \alpha/4\pi+\Order(m)$.
With a cut only on the photon energy, we find
\begin{align}
I_1^{\rm hard} &= -1+\epsilon-\log\epsilon + \Order(\delta) ~,\nn
I_2^{\rm hard} &= 2\Li(\epsilon)-\frac{\pi^2}{3}+2\log\delta\log\epsilon + \Order(\delta) ~,\nn
I_3^{\rm hard} &= -(1-\epsilon)\big[1+\log\delta-\log(1-\epsilon)\big] + \Order(\delta) ~,\nn
I_4^{\rm hard} &=  - \frac{1-\epsilon}{2} \bigg[\frac{5+3\epsilon}{2}+(1+\epsilon)\big(\log\delta-\log(1-\epsilon)\big) \bigg] \nn
&\quad + \Order(\delta) ~,
\end{align}
with $\epsilon = 2\Em/\sqrt{s}$ and $\delta=m^2/s$ as in the main text.
Imposing an additional cut on $\cos\theta_{e^\pm\gamma}=c_{e\gamma}$, we find in the massless limit $m=0$
\begin{align}
I_1^{e\gamma} &= 0 ~, \quad 
I_2^{e\gamma} = 2\log\epsilon \log\frac{1-c_{e\gamma}}{1+c_{e\gamma}} + \frac{\pi^2}{3} - 2\Li(\epsilon) ~,\nn
I_3^{e\gamma} &= (1-\epsilon) \bigg[1-\log(1-\epsilon) - \log\frac{1-c_{e\gamma}}{1+c_{e\gamma}}\bigg] ~,\nn
I_4^{e\gamma} &= -\frac{1-\epsilon}{2} \bigg[ 
\frac{3}{2}(1-\epsilon) + \frac{8}{1-c_{e\gamma}} \nn
& \quad \times \bigg( \frac{1+c_{e\gamma}}{(1-c_{e\gamma})(1-\epsilon)}
\log\Big(\epsilon+\frac{2(1-\epsilon)}{1+c_{e\gamma}}\Big) -1\bigg) \nn
& + (1+\epsilon)\bigg(\log(1-\epsilon)
+\log\frac{1-c_{e\gamma}}{1+c_{e\gamma}}\bigg) \bigg] ~. 
\end{align}
Finally, employing a cut on $\cos\theta_{e^+e^-}=c_{ee}$, the hard, non-singular part of the bremsstrahlung
integrals in the massless limit $m=0$ is
\begin{align}
I_1^{ee} &= 0 ~, \quad 
I_2^{ee} = - \Li\Big(\frac{c_{ee}-1}{c_{ee}+1}\Big) ~, \nn
I_3^{ee} &= -\frac{2}{1 - c_{ee}}\log\frac{1+c_{ee}}{2} -1 ~, \nn
I_4^{ee} &= \frac{1}{1 - c_{ee}} \bigg[ \frac{2c_{ee}}{1 - c_{ee}} \log\frac{1+c_{ee}}{2} + \frac{1+3c_{ee}}{4}\bigg] ~.
\end{align}

For completeness, we also add the equivalent of Eq.~\eqref{eq:bremsTraces} 
for the $P\to V\ell^+\ell^-$ and $V\to P\ell^+\ell^-$ decays discussed in Sect.~\ref{sec:other}.
The bremsstrahlung contribution to the correction factor can be written 
in the same form as Eq.~\eqref{eq:bremsOmega} with $N_i \to N'_i$, $i=1,2,3$, 
$\lambda \doteq \lambda(M_A^2,s,M_B^2)$, 
\begin{align}
N'_1 &= -4 \bigg\{ m^2 \Big[ \Big(1+\frac{4m^2}{s}\Big)\lambda + \nu_1^2 \Big] \nn[-1mm]
& \qquad 
+ \left(s_1-m^2\right)\left(M_A^2-M_B^2\right)\nu_2 \bigg\} ~,\nn 
N'_2 &= -4 \bigg\{ m^2 \Big[ \Big(1+\frac{4m^2}{s}\Big)\lambda + \nu_2^2 \Big] \nn[-1mm]
& \qquad - \left(s_2-m^2\right)\left(M_A^2-M_B^2\right)\nu_1 \bigg\} ~, \nn 
N'_3 &= 4 \bigg\{ (s_3-2m^2)  \Big(1+\frac{4m^2}{s}\Big)\lambda 
+ \frac{s}{2}\left(\nu_1^2+\nu_2^2\right)  \nn
& \qquad -2m^2\nu_1\nu_2 + \frac{1}{2s}\left[\lambda + \left(M_A^2-M_B^2\right)^2\right] \nn
& \qquad\quad \times \left[\left(s_1-m^2\right)^2+\left(s_2-m^2\right)\right] \bigg\} ~. 
\end{align}
The angular integration Eq.~\eqref{eq:angint} holds as before with the replacements 
$M_K\to M_A$, $M_\pi\to M_B$.
As already pointed out in Sect.~\ref{sec:other}, the result Eqs.~\eqref{eq:bremsIntComb} and \eqref{eq:defIi}
are then reproduced identically as in the decay $K\to\pi\ell^+\ell^-$.
\end{sloppypar}

\end{appendix}


\end{document}